\documentclass[10pt,journal,compsoc]{IEEEtran}

\usepackage{graphicx}
\usepackage{amsmath}
\usepackage{lineno,hyperref}
\usepackage{tabularx}
\usepackage{array}
\usepackage{tabularx}

\usepackage{booktabs} 

\usepackage[ruled]{algorithm2e} 
\usepackage{algorithmic}

\SetAlFnt{\small}
\SetAlCapFnt{\small}
\SetAlCapNameFnt{\small}
\SetAlCapHSkip{0pt}
\IncMargin{-\parindent}

\ifCLASSOPTIONcompsoc
  \usepackage[nocompress]{cite}
\else
  \usepackage{cite}
\fi

\begin{document}

\title{Privacy-preserving Health Data Sharing for Medical Cyber-Physical Systems}

\author{Han~QIU,~\IEEEmembership{Member,~IEEE,}
        Meikang~QIU,~\IEEEmembership{Senior Member,~IEEE,}
        Meiqin~LIU,
        Gerard~MEMMI,
\IEEEcompsocitemizethanks{\IEEEcompsocthanksitem Han QIU and Gerard MEMMI are with department of LTCI, Telecom-ParisTech, 75013, Paris, France. Meikang QIU is with department of Electrical Engineering, Columbia University, 10027, New York, USA. Meiqin LIU is with Zhejiang University, 310007, Zhejiang, China\protect\\

}
}


\IEEEtitleabstractindextext{%
\begin{abstract}
The recent spades of cyber security attacks have compromised end users' data safety and privacy in Medical Cyber-Physical Systems (MCPS). 
Traditional standard encryption algorithms for data protection are designed based on a viewpoint of system architecture rather than a viewpoint of end users. 
As such encryption algorithms are transferring the protection on the data to the protection on the keys, data safety and privacy will be compromised once the key is exposed. 
In this paper, we propose a secure data storage and sharing method consisted by a selective encryption algorithm combined with fragmentation and dispersion to protect the data safety and privacy even when both transmission media (e.g. cloud servers) and keys are compromised. 
This method is based on a user-centric design that protects the data on a trusted device such as end user's smartphone and lets the end user to control the access for data sharing. 
We also evaluate the performance of the algorithm on a smartphone platform to prove the efficiency.

\end{abstract}

}

\maketitle

\IEEEdisplaynontitleabstractindextext

%
\IEEEpeerreviewmaketitle

\section{Introduction}
\label{sec:intro}

{\em Cyber-Physical System} (CPS) is an integration of computation, including the cyber world (i.e. computers and the Internet) and the physical processes through computer networks. 
One important use case is a distinct class of CPS deployed in medical usage that combines smart medical devices, embedded software, and networking capabilities such as cloud computing which is referred as {\em Medical Cyber-Physical System} (MCPS)~\cite{lee2010medical}.

\begin{figure*}
\centering
\includegraphics[width=0.8\textwidth]{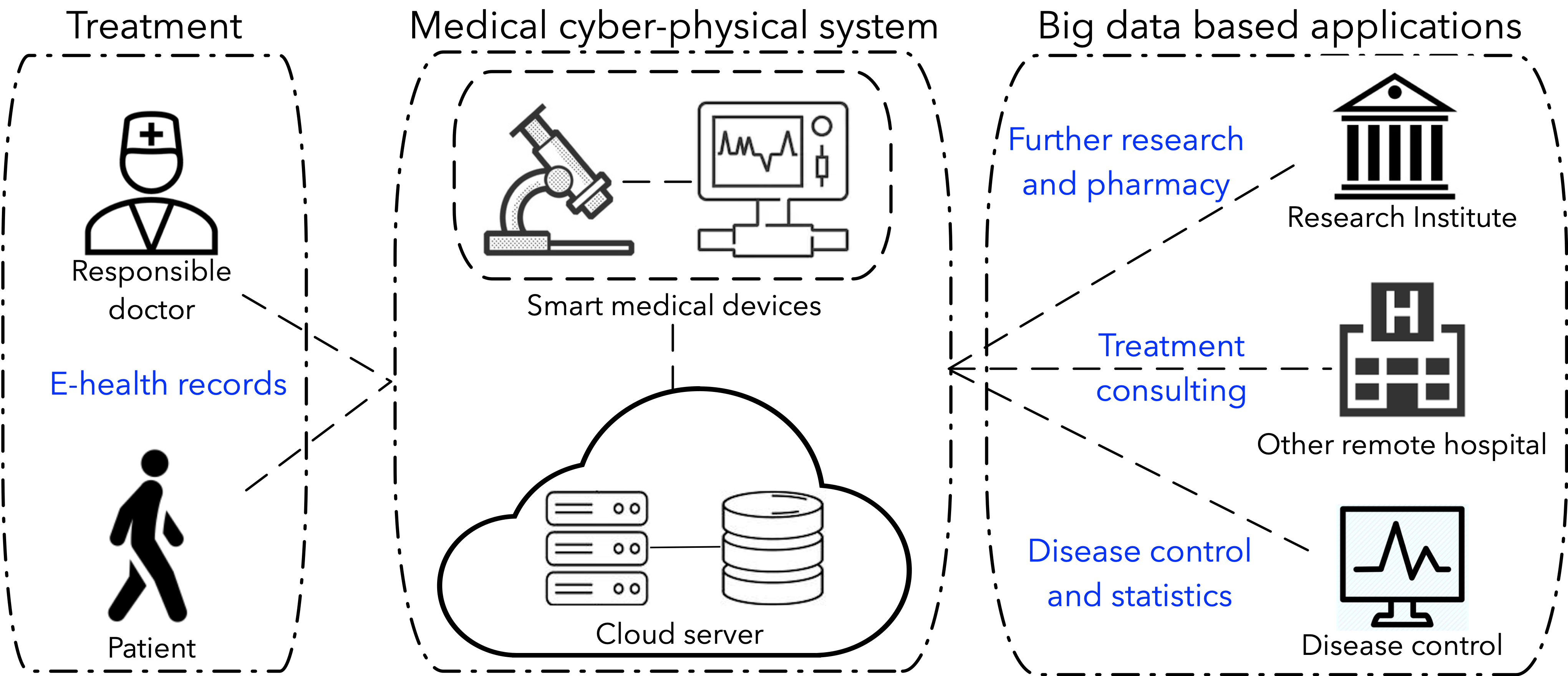}
\caption{Medical Cyber-Physical System (MCPS) use cases and applications based on big data.}
\label{fig:mcps1}
\end{figure*}

As shown in~\figurename~\ref{fig:mcps1}, in a cloud-based MCPS, the {\em Electronic Health Record} (EHR)~\cite{west2012mobile} can be stored and transmitted through cloud servers for other big data based applications. 
Collecting EHR data can not only help patients get better treatment from better hospital remotely but also provide inputs for research institutes for further pharmacy or study. 
Real-time health data monitoring by the disease control department can also improve the efficiency to prevent spreadable diseases. 


As EHR data is always sensitive, security and safety must be concerned when building such data sharing and processing system~\cite{act1996health}. 
Existing proposals for MCPS~\cite{zhang2017health} are more system-oriented rather than user-centric. 
For instance, the data preprocessing and protection scheme are deployed on clouds or third party servers that may involve plain data transmission on insecure Internet channels. 
Cloud service providers may also be unreliable due to the following two reasons. 
On one hand, any third party server may be curious as data privacy may be violated, for profiting by commercial purpose such as data mining based advertising. 
On the other hand, user behavior reports~\cite{stanton2005analysis} showed that end users may use the repeated or similar keys everywhere leading to potential dictionary attacks to their EHR data on clouds once keys are leaked from other sources.

Thus, with the preliminary that the cloud server cannot be totally trusted and the keys are vulnerable due to many practical reasons, we aim to provide EHR data safety and privacy from a user-centric viewpoint. 
More specifically, a design considering key reusing or exposure is required which means more protection methods must be introduced except only standard encryption algorithms. 
For instance, traditional data protection methods are basically key distributions or encryption algorithms which requires strong and different keys all the time. 
However, considering user behaviors pointed out in~\cite{stanton2005analysis} that most keys are just simple and repeatedly used in practical. 
Therefore, additional scheme must be introduced to guarantee the data safety on cloud ends even the keys are compromised. 
Our design is based on the assumption that the physical device of an end user such as a smartphone is trustworthy. 
Concepts like {\em Mobile Edge Computing} (MEC) allows data protection done on this smartphone before outsourcing. 
Then, the access control model must be designed and defined to let the end user to determine which entity can share the EHR data. 
Moreover, the proposed scheme must be able to provide the data safety and privacy under the situation that both cloud server and key are compromised 



Our contributions in this paper include: (1) we design a user-centric data storage and sharing method in cloud-based MCPS to protect the safety and privacy of users' EHR data which could protect data safety and privacy even when both cloud server and keys are compromised. (2) We evaluate the feasibility for this system based on mobile edge computing on a smartphone scenario to prove the improvement on efficiency compared with standard encryption algorithms.

We organize this paper following the order below. 
Section \ref{sec:issue} discuss the safety and privacy issue in MCPS by giving an example.
Section \ref{sec:design} illustrates our proposed user-centric EHR data protection system.
Section \ref{sec:analysis} provides the protection level of the selective encryption algorithm. 
Next, Section \ref{sec:evaluation} briefly evaluates the performance on a smartphone platform. 
Finally, Section \ref{sec:coc} draws the future work and the conclusion of this study.

\section{Research background}\label{sec:issue}

In this section, we briefly discuss the practical issue for threats over EHR data based on a user's viewpoint. 
An example is given to illustrate the threats we are facing in a simplified use case. 
Countermeasures and limitations in this special MCPS environment are also presented.

\subsection{Problem definition and threat models}


As pointed in Section~\ref{sec:intro}, on one hand, introducing cloud service for MCPS brings efficient, reliable and economic data storage and sharing for both patients and medical institutes, however, on the other hand, EHR data safety and privacy are challenged as well. 

\begin{figure}[!htbp]
\centering
\includegraphics[width=0.48\textwidth]{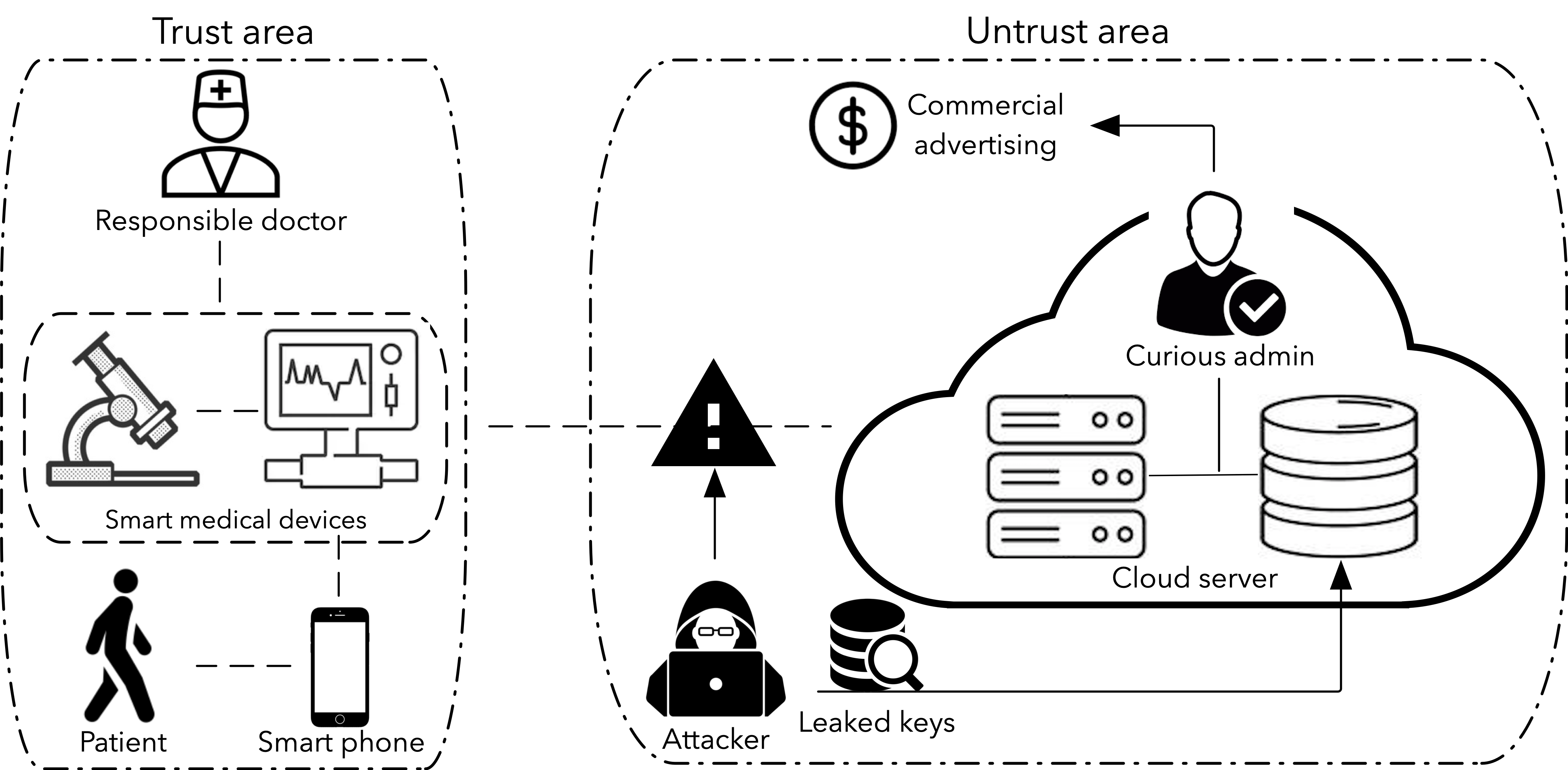}
\caption{A brief example for threats modes of EHR data on cloud based MPCS.}
\label{fig:example}
\end{figure}

In this example, we list several threat modes in~\figurename~\ref{fig:example}. 
The cloud based MCPS is abstracted as two areas and several components. 
Based on an end user's viewpoint, the responsible doctor and the personal device such as smartphones can be trusted. 
Once the health data is outsourcing to other parties in the cloud based MCPS network such as cloud servers, multiple threats will appear.

First, there are attackers trying to eavesdrop data on the public communication channel which may compromise data transmitted in plain text. 
Then, due to the fact that repeated keys are commonly used by end users, attackers may break into the cloud storage by using brute force attacks based on key dictionary collected from other leakages~\cite{o2017giant, jingdong2016}.
Moreover, the cloud service providers may also try to mine user's data to get profit such as data based advertising which violates privacy~\cite{stanton2005analysis}.
The worse case is that the cloud service providers may directly hand over plain stored data for authority's illegal surveillance without noticing the users~\cite{ball2013nsa}.



Protecting EHR data before outsourcing is necessary to avoid directly eavesdropping over communication channel or any operations on the cloud ends to compromise end users' data safety. 
However, requiring end users to set strong and different keys is a system-centric design, and more importantly, can not fundamentally solve the threats from attackers with dictionaries. 
Thus, the design must be able to avoid data leakage in the worst case assuming that an attacker has data stolen from cloud servers and the repeated key leaked from other sources.

\subsection{Research motivation}

As shown in~\figurename~\ref{fig:example}, the threats come from multiple parties in the MCPS network. 
The countermeasure is to design a user-centric data protection system that protects EHR data in a limited trust area and see all other parties as untrusted~\cite{qiu2017efficient}.

Thus, the data could still suffer from the leakage of other communication entities when there are malicious users in this environment expose the keys. 

In this paper, our motivation is to provide safety and privacy in the cloud-based MCPS against user behaviors such as repeated key usages through untrusted cloud servers. 
Fragmentation will be introduced to be combined with encryption such that the fragmented data pieces on clouds cannot be used to leak the stored data even when the key for this system is leaked. 
Access control model is also needed for data sharing with security and privacy, and further, especially for disease control in this MCPS network. 

We first define the very basic assumptions based on end user's viewpoint.  
The trusted area includes the smart medical devices equipped with processing and communication modules which forms the edge of the MCPS network. 
Another reasonable assumption is that every user has a trusted smartphone within the cloud based MCPS network that can perform communication and calculation tasks. 
Responsible doctor, introduced as the other trust party in MCPS network, must be also considered to be able to access the plain data for disease control purposes. 

Other assumptions are that the Internet transmission channel cannot be trusted so data outsourced must deploy protection scheme. 
The cloud service providers cannot be entirely trusted. We have to assume that one "curious" or "malicious" program sits on one cloud server and is able to observe all the data stored in the Cloud and transmitted through the Cloud. 
In a worse case, not only data stored on the cloud server but also the key for protection can be used by this program to sniff the users' privacy by any means of analysis or attack for even a small piece of data. 
The purpose of this design is to simplify the basic roles in the complicated cloud based MCPS network. 
Starting from a user-centric viewpoint is to use the big data and cloud based MCPS for better medical treatment purpose while also protect each user's safety and privacy.

\section{Design and implementation}\label{sec:design}

In this section, we present the design of user-centric secure data sharing system. 
First, system architecture is given to illustrate the idea that protecting data by fragmentation to avoid direct relationships between keys and data fragments stored on clouds.
A SE algorithm is given to support this design. 
Access control model is also given considering the specialty of data sharing in MCPS.

\subsection{System designs}

The initial idea is to protect data in a trusted place and then to fragment the data into fragments with different levels of importance. 
A dispersion step will be followed to determine which data fragments will be stored on cloud servers. 
The SE algorithm guarantees that only the keys cannot be used to decrypt the fragments stored on cloud servers.

\begin{figure*}
\centering
\includegraphics[width=0.9\textwidth]{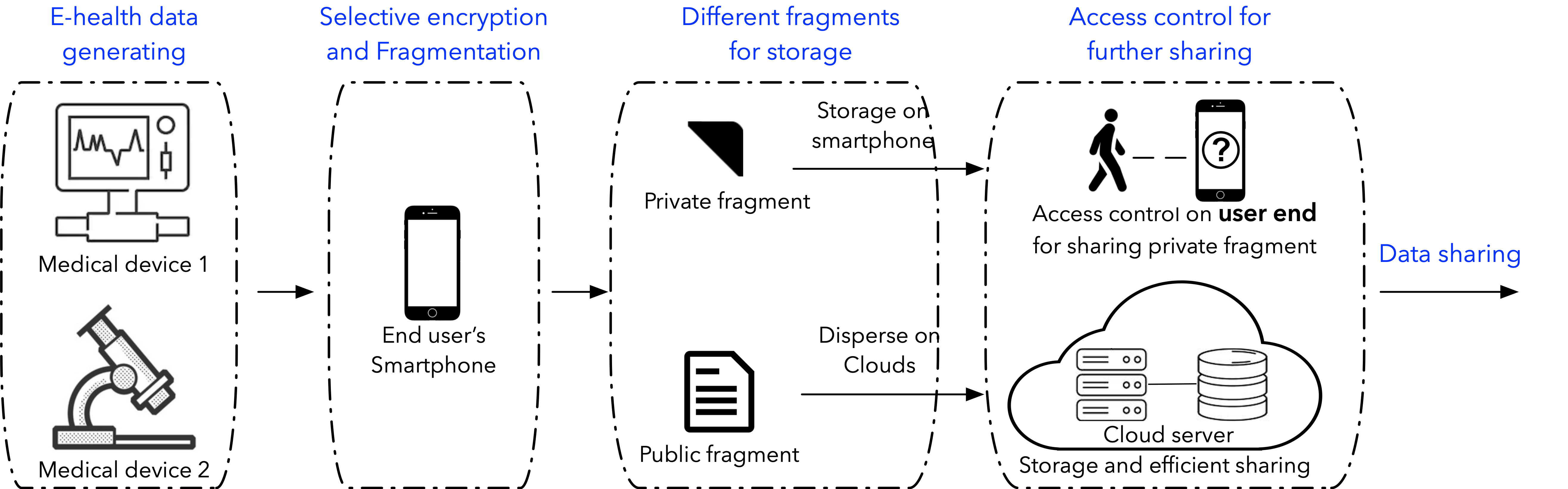}
\caption{System architecture based on user-centric viewpoint.}
\label{fig:system1}
\end{figure*}

In the MCPS practical environment, we could assume that the biological or medical information is generated on smart medical devices which are formatted EHR files as shown in Step 1 in~\figurename~\ref{fig:system1}. 
This health data can be shown on the medical devices for treatment by doctors and also be transmitted to the patient's personal devices such as smartphones in this paper. 
A direct and offline connection can be built between the medical devices and smartphones for the data transmission as the Step 2. 
In this step, the smartphones will receive the health data with the modification from doctors as the input. 
SE methods then will be deployed on the smartphones to fragment and selectively encrypt the fragmented data. 
Considering the limitation of storage and transmission of smartphones, data piece with low importance level (called {\em public fragment} as shown in~\figurename~\ref{fig:system1}) will be transmitted to cloud servers and the important data piece (called {\em private fragment} as shown in~\figurename~\ref{fig:system1}) will be stored on the smartphones in this scenario. 

The SE method and the fragmentation scheme must be able to guarantee that the private fragment is small in data size but contains most of the important information to recover the initial plain data. 
While the public fragment must be large in data size but cannot be used to recover the initial data no matter by attacker or administrators in cloud servers as it is protected and not related with the keys.
For a further data protection, the private fragment should be encrypted and backup in a cloud server with higher trust level in case the loss of smartphones. 
The secure storage can be achieved based on an end user's viewpoint that the private fragment is carried and there are no worries for any data leak at the cloud end.

For the EHR data sharing, the end user can always determine first whether to share with medical institute by control the data sharing of private fragment just on smartphones as Step 5 shown in~\figurename~\ref{fig:system1}. 
Once institutes are authenticated with access to one patient's health data, the public fragment can be downloaded from the cloud server and private fragment can be shared anonymously (remove the sensitive information such as identity). 
Thus, the cloud server, as an efficient data storage and sharing middleware, is fully deployed and data leaks can be avoided.

\subsection{General data protection methods}

EHRs data may include a large range of data types with different formats, including static digital data such as personal statistics like age and weight, graphics data such as demographics and radiology images, and data formats used for medical software such as laboratory test results or dynamic health statistics. 
Based on a user-centric viewpoint, if there are too many options to protect for choosing, it may cause reluctant to use~\cite{stanton2005analysis}.
Therefore, one general protection method must be designed to let different digital files with different formats protected in a simple way. 

Traditional encryption algorithms such as AES~\cite{daemen2013design} could be one choice. 
However, management of the key will be a problem as the security property is all relying on the storage of the key.
Assume one scenario that an end user encrypts all the EHRs with AES and upload the ciphertext to the cloud servers, data could be leaked once the key is compromised. 
According to the user behaviour~\cite{das2014tangled}, many end users might reuse the same keys in different places such that one leakage of the keys could lead to serious leakage of privacy data elsewhere. 
Fragmentation is introduced such that the attackers cannot crack the data content even the key is compromised. 
One approach is to use lossless transformations (DWT) to fragment one digital file into different fragments with different importance levels~\cite{qiu2017efficient}. 
However, this design is not suitable for protecting EHRs as EHRs are a series of many different files instead of large data chunks. 
In summary, a more practical and deployable method is needed to protect the EHRs in a manner that can protect many different files in a general way and compromising key is not enough to threaten the safety and privacy of the data.

\subsection{Selective encryption algorithm}

The basic idea is to fragment the digital data in a manner that make different data fragments related. 
For instance, a small subset of the data is used to protect the rest data fragment in a lightweight manner. 
Then protection scheme such as encryption algorithms can be used to protect the small subset of the data with a key. 
A dispersion scheme is used for the storage that the encrypted small subset of the data is stored in a secure place such as end user's personal device and the rest data fragments are stored in cloud servers for cost saving purposes. 
In such case, once the key is leaked, even the attacker can get access to the data stored on cloud servers, the data content is still safe as the key cannot be used to decrypt the fragments stored on clouds. 
By this design, the data leakage due to the password reuse~\cite{das2014tangled} can be avoided even the encryption key is also leaked.

In this section, we introduce the algorithm to selectively encrypt the data and provide a dispersion method for further storage. 
The setting shown in~\tablename~\ref{tab:notations} for the parameters is used for fitting different file formats of EHR data. 
Firstly the EHR data input $D_{input}$ (image, database file, etc) is preprocessed as the file header with all necessary markers for format $D_{head}$ and the content $C$ which is $D_{input} = D_{head} + C$.
The size of $D_{head}$ is normally ignorable compared with $C$ and it is supposed to be stored in plaintext locally. 
Then the $C$ is the input for the SE algorithm with a key $K$ and a $Counter = 0$. 
The key generation and distribution process are the same with the standard protocols such as SSL/TLS protocols. 
The outputs are the corresponding private fragment $PRF_{c}$ and public fragment $PUF_{c}$.


\begin{table}[!htbp]
\caption{Major notations used in this algorithm and their definitions.}
\label{tab:notations}
\begin{tabular}{l l }
\hline
Notation                  & Definition  \\ \hline
$D_{input}$               & Input EHR data files       \\ 
$D_{head}$                & File headers containing markers such as formats       \\ 
$C$                       & EHR data contents       \\ 
$K$                       & Secret key used to protect the data      \\ 
$Counter$                 & an integer for counting processing sequence        \\ 
$PUF_{c}$                 & Corresponding public fragment          \\ 
$PRF_{c}$                 & Corresponding private fragment          \\\hline
\end{tabular}
\end{table}


As in the algorithm, each file content $C$ is processed as many 256-bit units, even very small EHR files can be processed independently which can further reduce the modification times if only some small EHR files are modified. 
The basic idea is to use the keys to protect the important part which is the private fragment. 
Then the fragment combined with the keys will be used to protect the public fragment with an approximated one-time pad protection. 


More details are listed in the Algorithm 1. 
We initialize the algorithm in Line 1-3 to set all parameters and also read input EHR data content $C$ as $N$ units with each unit of 256-bit size: $C = \{C_{1}, C_{2}, ... , C_{N}\}$. 
Then each data unit $C_{i}$ will be firstly fragmented into 8 fragments of 32 bit size ($C_{i} = \{C_{i}^{1}, C_{i}^{2}, ... , C_{i}^{8}\}$) in Line 5. 
In Line 6, a pseudorandom number generator will be used to determine which fragment will be the private fragment ($C_{i}^{j}$) and the rest 7 fragments ($C_{i} - C_{i}^{j}$) will form the public fragment for this 512 bit data unit. 
In Line 7-9, the protection for the public fragment is done by Xoring the $C_{i} - C_{i}^{j}$ with the Hash results of $C_{i}^{j}$, $K$, and $Counter$. 
The processing for the sequence of $C$ is done in a for loop in this Algorithm. 
However, if the pseudorandom number sequence $j$ is pre-calculated, the processing for the sequence of $C$ can be then parallelized  for better performance.

In this algorithm, the SHA-256 function~\cite{gueron2011sha} is introduced to generate bit sequences with a 50\% difference even when there is only 1 bit changed in the input. 
Thus, bit sequence used to protect the $PUF_{c}$ by xoring is unpredictable which can be seen as the approximated one-time pad protection~\cite{croft2005using}. 
The decryption process of this algorithm is symmetric to the encryption algorithm.
Data recovery can only be achieved when both private and public fragments are collected with the correct keys. 
More security analysis will be introduced in the Section~\ref{sec:analysis} to prove the protection results.


\begin{algorithm}
\caption{Selective encryption algorithm with fragmentation}
 {\bf Input:} Data $D_{input}$ and Key $K$.\\
 {\bf Output:} Public fragment $PUF_{c}$ and private fragment $PRF_{c}$.\\
\begin{algorithmic}[1]
 \STATE$D_{input} = D_{head} + C;$/*Protection on only the data content.*/\\
 \STATE$C = \{C_{1}, C_{2}, ... , C_{N}\};$ /*Read input data $C$ as $N$ units with each unit $C_{i}$ size as 256 bits*/\\
 \STATE$PRF_{c} = \{\};$$PUF_{c} = \{\};$/*Initialize the public and private fragments as null sequence sets*/\\
 \STATE$Counter = 0;$\\
 \STATE{\bf for} $i \gets 1$ to $N$ {\bf do}
  \STATE$C_{i} = \{C_{i}^{1}, C_{i}^{2}, ... , C_{i}^{8}\};$ /*Fragment $C_{i}$ equally into 8 sequences with each sequence $C_{i}^{j}$ size as 32 bits*/\\
  \STATE$j = PRNG (K);$ /*$j$ is an integer from a pseudorandom number generator with key as a seed and $j\in[1,8]$*/\\
  \STATE$H_{i} = SHA_{256}(C_{i}^{j} + K + Counter);$ /*$H_{i}$ is a 256-bit sequence generated by Hashing $C_{i}^{j}$, $K$, and $Counter$*/\\
  \STATE$PRF_{c} = Add (PRF_{c}, C_{i}^{j});$ /*$C_{i}^{j}$ is collected into the private fragment $PRF$*/\\
  \STATE$PUF_{c}^{i} = XOR(C_{i} - C_{i}^{j}, truncate(H_{i}));$ /*Public fragment $PUF_{c}^{i}$ for $C_{i}$ is protected by xoring the 7 rest 64-bit sequences with the truncated $H_{i}$*/\\ 
  \STATE$Counter = Counter + 1;$ /*Counter updated every loop to guarantee different $H_{i}$ even $C_{i}s$ are the same*/\\
 \STATE{\bf end for}\\
 \STATE$PRF_{c} = ENC_{AES}(PRF_{c}, K);$ /*Private fragment $PRF_{c}$ is protected by AES with key $K$*/\\
 \STATE{\bf Return} $PRF_{c}$, $PUF_{c};$
 \end{algorithmic}
\end{algorithm}

\subsection{Mobile edge computing}

In traditional designs, MEC technology is usually used to offload calculations from the clouds end which can further avoid the data transmission between edge devices and cloud servers~\cite{liu2017scalable}. 
Typical edge devices such as smartphones are equipped with processing cores but limited storage spaces which fits the needs for deploying MEC to process data and upload data to cloud servers. 
In this paper, we claim that in some scenarios, MEC designs can also enhance the security level for personal data. 
For instance, as long as the data is outsourcing in plaintext to the communication channel, potential attackers might get the data by eavesdropping as shown in~\figurename~\ref{fig:example}. 
Data leakages as shown in~\cite{ball2013nsa} and in Facebook~\cite{facebook2018} recently have proven that totally trusting cloud service providers is not wise.

Thus, as the smartphones are normally physically in each end user's hand, the basic assumption is that the end user's digital device can be trusted. 
The protection on smartphones then outsourcing data can protect sensitive data from potential attackers and data leakage at cloud ends. 
Of course, {\em Fully Homomorphic Encryption} (FHE)~\cite{gai2017blend} is promising to provide full privacy for storage or calculation but it seems the performance issues are still the obstacles in practical. 
Thus, in this paper, we claim that the MEC can bring safety and privacy for end user's data as long as the protection method is implemented on smartphones before outsourcing.




\section{Protection analysis}\label{sec:analysis}

In this section, we test the protection levels for several typical EHR files. 
As mentioned in Section~\ref{sec:design}, the private fragment is protected by AES and a public fragment is protected by the SE algorithm. 
First, we assume the AES encryption algorithm is secure (AES is used as the standard encryption algorithm in this paper but can be replaced with other encryption algorithms).  
Then the threat model is that the attackers may have access to the public fragments. 
Thus, we need to prove that the public fragments cannot be used to recover the original data content. 
In order to meet this requirement, the public fragments should be similar to random values no matter what the input file format is. 

\subsection{Statistical analysis on the public and protected fragments}

\begin{figure}[!htbp]
\centering
\includegraphics[width=0.48\textwidth]{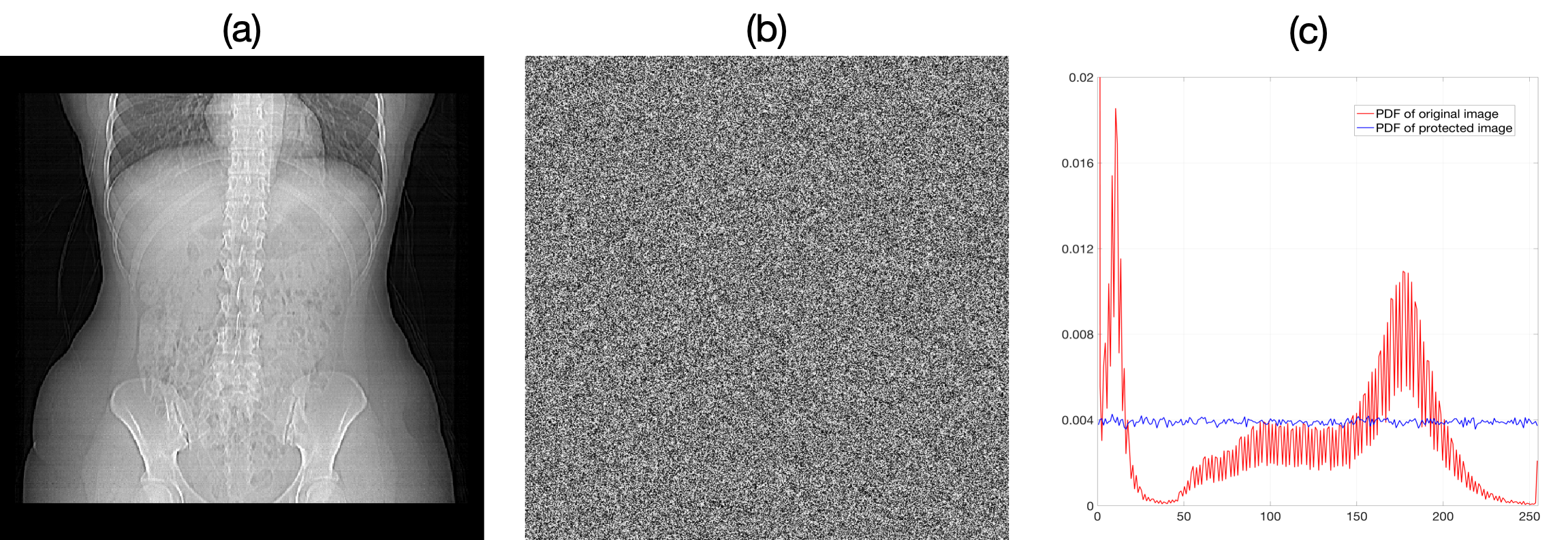}
\caption{(a) Original DICOM image, (b) protected DICOM image, and (c) their PDFs.}
\label{fig:pdf1}
\end{figure}

In this analysis, two main criteria are used to test that the randomness level compared with the input EHR data and the output public fragments. 
Firstly, we test one typical medical image in {\em Digital Imaging and Communications in Medicine }DICOM format~\cite{mildenberger2002introduction}. 
This image format is common to be used in medical image storage and transmission. 
As shown in~\figurename~\ref{fig:pdf1}, we present the input DICOM image (a) and the corresponding public fragment (b). 
A {\em Probability Density Function} (PDF) analysis is given as well to indicate that the byte value distribution of the public fragment is nearly random. 
Then according to our dispersion design, this public fragment is supposed to be stored on cloud servers. 
Even the cloud server is somehow compromised and data is leaked, this public fragment file cannot be used to recover the original DICOM image as the byte values in it are uniformly distributed which can be seen as nearly random.

As there are actually many different file types of EHR, we test some typical file formats to prove that the designed protection scheme can always generate public fragments that are similar to random data. 
One important criteria is to measure the entropy values for the input EHR data and the corresponding public fragments. 
Entropy values of a data sequence is a parameter that measures the level of uncertainty in a random variable~\cite{zhang2011novel}.
In our test, the entropy values are used to prove the high randomness level of the public fragments. 
Thus, we calculate the entropy values for the input EHR data and also the corresponding public fragments. 
It is easy to calculate for a random source emitting $2^{N}$ symbols, the entropy should be $N$. In this design, as the data are always seen as 8-bit per element, the pixel data have $2^{8}$ possible values. As such, the entropy for a "true random" information source must be 8. For our case, the entropy of the public fragments is always more than 7.99 which proves high randomness. 
In~\tablename~\ref{tab:entropy}, we test also two EHR data samples downloaded from ONC website~\cite{hitdata}. 
According to the entropy values, the high level of randomness is achieved compared with the input EHR database format files.

In summary, with the different tests for some other different EHR file formats, we get similar results that the high randomness level is achieved for the public fragments. 
Thus, according to the threat mode described in Section~\ref{sec:issue}, neither the attacker nor the cloud admins cannot compromise the EHR data content even with full access to the public fragments on clouds.

\begin{table}[!htbp]
\caption{Entropy tests for the protected public fragments of DICOM images and EHR data samples.}
\label{tab:entropy}
\centering
\begin{tabular}{l l l}

\hline
Notation                  & Entropy of input files & Entropy of $PUF_{c}$  \\ \hline
DICOM image 1                & \multicolumn{1}{c}{6.402} &\multicolumn{1}{c}{7.9992}       \\ 
DICOM image 2                & \multicolumn{1}{c}{6.307}  & \multicolumn{1}{c}{7.9993}       \\ 
EHR data sample 1            & \multicolumn{1}{c}{4.887} & \multicolumn{1}{c}{7.9970}      \\ 
EHR data sample 2            & \multicolumn{1}{c}{4.992} & \multicolumn{1}{c}{7.9987}         \\\hline
\end{tabular}
\end{table}

\subsection{Privacy-preserving data sharing with user-centric designs}

\begin{figure*}
\centering
\includegraphics[width=0.9\textwidth]{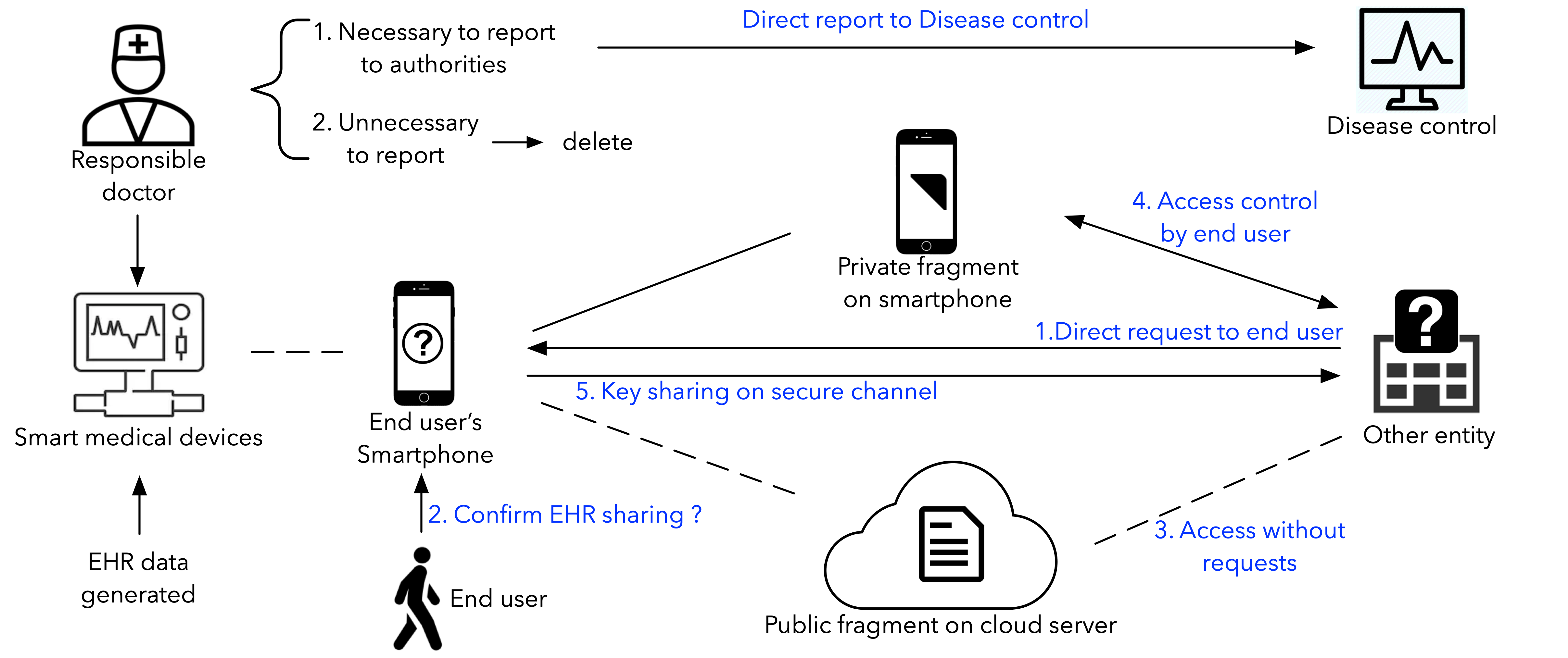}
\caption{EHR data sharing model with access control from an end user.}
\label{fig:access1}
\end{figure*}

Once the EHRs are protected and dispersed to different storage space, the secure storage step is finished. 
For further data sharing, the data sharing protocol is shown in~\figurename~\ref{fig:access1}. 

Due to the specialty of the medical system, the responsible doctor should have a higher access control role in this protocol to guarantee serious conditions like spreadable disease should be reported in the first time. 
As shown in Fig. 4, the responsible doctor can directly access the medical data with the smart medical devices. 
The scheme proposed in this paper is used to protect the EHR data when the data needs to be shared between other parties through untrusted public clouds. 
Thus, the delay caused by the security functions will only affect on the communication between other parties such as remote hospitals. 
Then the end user should have the private and public fragments for the EHR data stored on the smartphone and cloud servers respectively. 
In this protocol, the trust entities are the doctors, disease control authority and the end user himself. 
The other entity who wants to get the EHR data is untrusted and can only get the public fragments without the authority from the end user.

One important user behavior is that the users are more likely to use the simple key and reuse the same key~\cite{stanton2005analysis}. 
The direct result by this behavior is that the attackers might easily guess the key which will let the full encryption protection useless. 
The other scenario is that once one database is compromised, attackers might try to compromise other databases with the dictionary formed by compromised keys which leads to more serious data leakages. 

Considering the threat models listed in Section 2.1, once the key used to encrypt the private fragment is exposed and acquired by a malicious party, the data is still protected as long as the smartphone is secure. 
Furthermore, the malicious might also get access to the public fragment stored on the cloud server. 
However, as the encryption key is used to protect the private fragment and the public fragment is protected based on the plain text of the private fragment, the data is still protected as the encryption key cannot be used to decrypt the public fragment. 
Considering the worse case that an end user is repeatedly using the same key for encryption and the public fragment is always vulnerable due to the cloud server is curious, the data is still protected as the private fragment is secure. 
Moreover, even if the attacker could pretend to be valid party for one time to get the private fragment for one EHR data file which would compromise this particular EHR data, the other EHR data files are still protected as the end user could still deny the access for the private fragments of the other EHR data files.

\subsection{Discussion}


In this work, we introduced the fragmentation and dispersion for outsourcing and storage for designing the SE algorithm. 
The basic idea is to first use the key to protect a small part of the data, and then to use this small part of the data to protect the rest large part of the data. 
In such case, the key is not directly related with the large part of the data which is supposed to outsourced on cloud servers. 
This could defend the data leakage due to reuse of password that is even the attackers know the key and have access to the public fragments, the data is still secure. 
However, there are other viewpoints such as the security level of smartphones are difficult to guarantee as a smartphone might get stolen or compromised by malicious applications. 
In fact, the security of operating systems on smartphones is not in the scope of this paper.

\section{Performance evaluation}\label{sec:evaluation}

The MCPS is a system that requires low latency especially for critical health conditions. Thus, for the EHR data sharing with remote communication entity, we evaluate the proposed algorithms on a smart phone platform to prove the efficiency of the proposed algorithm. 
The baseline model to be compared is to encrypt all EHR data on the smartphone end with AES-128 with CBC mode and then outsource the cipher texts. 
In the practical data sharing scenario as shown in~\figurename~\ref{fig:access1}, the actual latency of data sharing is consisted by many factors such as the data transmission latency, the latency of access control operation, and the latency caused by data protection. 
As we are proposing a novel data protection scheme, in this section, we only evaluate the execution time of the proposed algorithm compared with the standard encryption algorithm. 
For the other latency such as transmission latency, we assume that the latency for our proposal is the same with the scenario when standard encryption algorithm is deployed as we do not increase the data volume for transmission. 

Actually, according to the SE algorithm, there are only two heavy calculation tasks involved: AES encryption (assume as AES-128 with CBC mode) and Hash operation (SHA-256 in this paper).  
Firstly, the input data content $C$ is cut into many units with unit size 256 bits. 
There are 32 bits selected to form the private fragment and one SHA-256 operation will be performed for this unit.
Thus, for each input data content $C$, there will be 12.5\% (every 32 bits out of 256 bits) to be encrypted with AES-128 and there is only one Hash operation for every 256 bits which means there will be 32k Hash operations of SHA-256 for 1 MB EHR data.

\begin{figure}
\centering
\includegraphics[width=0.48\textwidth]{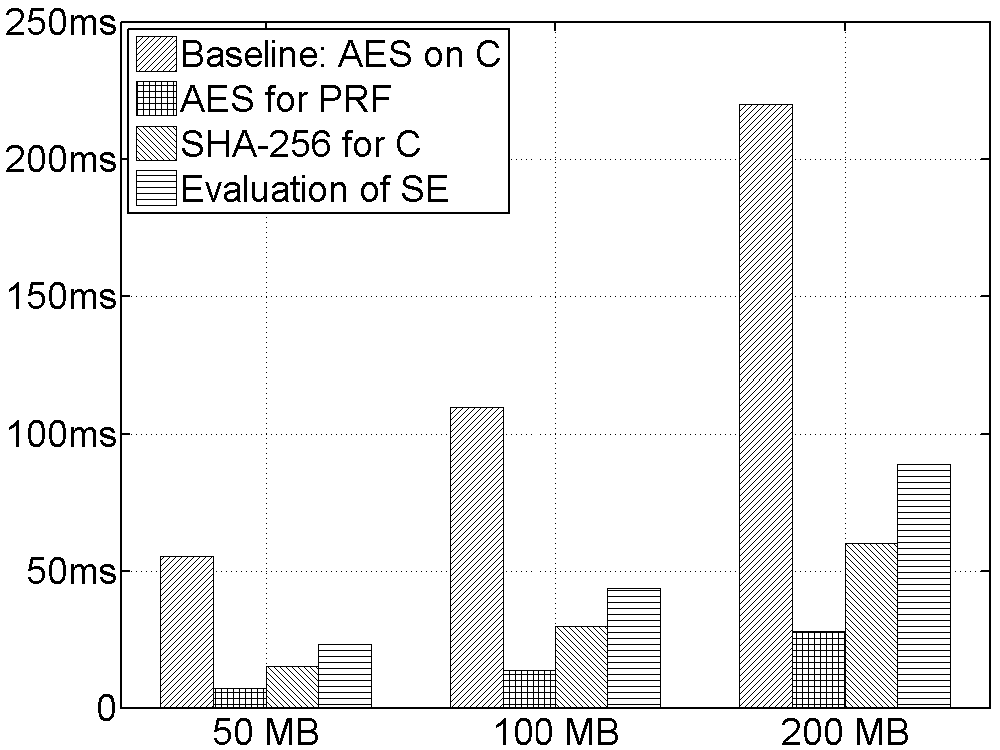}
\caption{Evaluated performance comparison between proposed SE algorithm with the AES-128.}
\label{fig:perf1}
\end{figure}

We evaluate the execution speeds for the main calculation tasks on iPhone 8 Plus which equips the new A11 processor which contains six ARMv8 CPU cores (2 high-performance 2.53 GHz and 4 high-efficiency 1.42 GHz)~\cite{jacobs2017apple}. 
Based on the experimentation in~\cite{wolfssl}, the A11 processor could enjoy the huge acceleration brought by the ARMv8 crypto extensions and single precision math. 
The AES-128 with CBC mode could achieve 912.347 MB/s and the SHA-256 performance is 1717.28 MB/s. 
Thus, the performance evaluated for the proposed algorithm is shown in~\figurename~\ref{fig:perf1}. 
We tested the evaluated time consumed by the baseline scenario which is the AES-128 with CBC mode for the data content $C$. 
Then, we test the time consumed by the AES (AES-128 with CBC mode) for the $PRF$ of $C$, the SHA-256 for the data content $C$, and the evaluated SE which is the total time consumed by the two main calculation tasks respectively. 
The evaluation result of the proposed SE algorithm on the iPhone 8 Plus platform with the acceleration of ARMv8 crypto extensions is 2.2GB/s which is approximately 2.3 times faster than the baseline scenario of AES-128 with CBC mode. 
Although the evaluation for the individual calculation tasks show that the execution time of this algorithm is much less than using full encryption with AES, there are still other factors to cause latency for the practical implementation. 
However, according to the test in this paper, we prove that on a smartphone platform, it is faster to use the proposed SE method than the full encryption.




\section{Future work and conclusion}\label{sec:coc}

For the future work, we will deploy this scheme with a more practical environment such as a multi-cloud scenario~\cite{kapusta2015data} and also discuss the possibility to define different cloud servers with different trust levels and disperse all fragments on clouds with either encryption or SE. 
Also, more implementation works to test the speed of AES and Hash on different smartphone platforms are also needed in the future.


In this paper, we presented an EHR data protection and sharing system for MCPS based a user-centric viewpoint. 
We aimed to solve the threats to data safety considering end users' behaviors that keys might be reused and leaked from other sources.
A SE algorithm combined with fragmentation and dispersion for storage is designed to protect data even when both the key and the public fragment of EHR data on clouds are leaked. 
The test for the protection results proved that our algorithm can avoid data recovery even with both key and public fragments compromised.

\section*{Acknowledgement}

This work was supported by the Joint Research Fund for Overseas Chinese, Hong Kong and Macao Scholars of NSFC under Grant 61728303 and the Open Research Project of the State Key Laboratory of Industrial Control Technology, Zhejiang University, China under Grant ICT170331.


\bibliographystyle{IEEEtran}
\bibliography{mybibfile}

\end{document}